# Ferromagnetism in substituted zinc oxide

M. Venkatesan, C. B. Fitzgerald, J. G. Lunney and J. M. D. Coey,

Physics Department, Trinity College, Dublin 2, Ireland.

***Abstract.*** Room-temperature ferromagnetism is observed in (110) oriented ZnO films containing 5 at % of Sc, Ti, V, Fe, Co or Ni, but not Cr, Mn or Cu ions. There are large moments, 1.9 and 0.5$\mu_B$/atom for Co- and Ti-substituted oxides, respectively. Sc-substituted ZnO shows also a moment of 0.3 $\mu_B$/Sc. Magnetization is very anisotropic, with variations of up to a factor three depending on the orientation of the applied field relative to the R-cut sapphire substrates. Results are interpreted in terms of a spin-split donor impurity band model, which can account for ferromagnetism in insulating or conducting high-*k* oxides with concentrations of magnetic ions that lie far below the percolation threshold. The variation of the ferromagnetism with oxygen pressure used during film growth is evidence of a link between ferromagnetism and defect concentration.

Following the original report by Ueda et al. [1], there have been sporadic but inconsistent reports that the wide-gap semiconductor ZnO exhibits ferromagnetism with a Curie point above room temperature when thin films of the oxide are doped with a few atomic percent of cobalt [2-6] or another transition element [7-10]. The results are sensitive to the form of the sample and preparation method. Other studies found lower magnetic ordering temperatures [11-14], or no ferromagnetism at all above 3 K for any $3d$ dopant [15]. In the absence of an exchange mechanism which could account for a high Curie temperature at doping levels far below the percolation threshold, these reports have been received with skepticism, and the belief that the ferromagnetism must somehow be associated with clustering or incipient formation of secondary phases. But there is spectroscopic evidence that divalent cobalt does indeed substitute on the tetrahedral sites of the wurtzite structure [1, 11,17], with a wide solid solubility range [15]. A search by Rode et al [4] revealed no evidence for phase segregation in Co-doped ZnO films. Nevertheless, until a clear connection between the magnetic properties and electronic structure can be shown, doubts that doped zinc oxide is truly a magnetic semiconductor will persist.

We recently proposed a model for high-temperature ferromagnetism in dilute $n$-type magnetic semiconductors, where the exchange is mediated by carriers in a spin-split impurity band derived from extended donor orbitals [18]. The model predicts ferromagnetism when the carrier concentration per cation δ exceeds $4.4/\gamma^3$, where γ = $k$m/m*; $k$ is the high-frequency dielectric constant and m* is the electron effective mass; the value of γ for ZnO is 14. To avoid antiferromagnetic coupling, the concentration of magnetic cations $x$ must lie below the cation nearest-neighbour percolation threshold $x_p$, which is ≈ 0.17 for the wurzite structure. High Curie temperatures are expected when $3d$ states of the transition element hybridize with the spin-split impurity-band states at the Fermi level, providing charge transfer from the impurity band to the $3d$ ion of $0.01 - 0.02$ electrons per cation.

In order to probe the origin of the high-temperature ferromagnetism in $(Zn_{1-x}M_x)O$, we have examined the systematic variation of the magnetism i) as a function of transition metal M at fixed concentration $x = 5\%$, ii) as a function of transition-metal concentration for M = Co and iii) as a function of oxygen pressure during film growth for $x = 5\%$ and M = Co.

All samples were prepared by conventional pulsed-laser deposition using a 20 ns, 248 nm KrF excimer laser at a fluence of 1.8 J cm$^{-2}$. Targets were prepared by standard ceramic procedures from ZnO and MO or $M_2O_3$ powders. The substrates were 5 x 5 mm$^2$ R-cut ($1\bar{1}02$) sapphire maintained at 600 °C. The oxygen pressure during deposition was varied from 1 to $10^{-4}$ mbar. Film thickness was in the range 60 - 120 nm. It was measured with a precision of ±10% using an in-situ optical reflectivity monitor during deposition and checked by small-angle X-ray reflectivity measurements. The films are transparent and strongly-textured, showing only the (110) reflection of the wurtzite structure. Film composition was determined by EDAX analysis for a series of 10 cobalt films prepared from targets with nominal composition x′ ranging from 0.01 to 0.15. The relation between x′ and x is almost linear, with x being about 30% greater than x′ [19].

Substitution of the cobalt cations in the tetrahedral sites of the wurtzite structure was confirmed by optical spectroscopy. A series of characteristic optical absorption bands in the infra-red and visible have been identified with $d$-$d$ transitions of the high-spin Co$^{2+}$ $3d^7$ $^4F$ ion in tetrahedral oxygen coordination [16,20]. The absorption in the visible was thought to derive from the $^4A_2 \to {}^4T_1$ ($^4P$) and $^4A_2 \to {}^2E$ ($^2G$) transitions, and those in the infrared from the $^4A_2 \to {}^4T_1$ ($^4F$) transition. The orbital triplet final states are split into a singlet and a doublet by the trigonal component of the crystal field at the tetrahedral site, which has 3*m* symmetry. The strength of these absorption bands scales with the cobalt content of our films [19].

The variation of the magnetic moment across the $(Zn_{0.95}M_{0.05})O$ series is illustrated in Fig.1. Measurements were made in a Quantum Design MPMS-XL SQUID magnetometer using linear regression and no auto-tracking. There is a striking systematic variation, with maxima near the beginning of the 3*d* series at Ti (0.5 $\mu_B$), and towards the end of the series at Co (1.9 $\mu_B$). No moment (< 0.1 $\mu_B$) was observed at room temperature for Cr, Mn, Cu or Zn. The other ions, including scandium, exhibit smaller moments. Ferromagnetism in $(Zn_{0.95}Sc_{0.05})O$ is quite a surprise, since neither $Zn^{2+}$ ($3d^{10}$) nor $Sc^{3+}$ ($3d^0$) are magnetic ions, and the 3d level of zinc lies well below that of scandium [21], so no electron transfer to scandium is envisaged.

Another surprise is the huge anisotropy of the magnetization. This has been measured for Sc, Ti, V, Fe, Co and Ni. Data for Co are presented in Figure 2. In every case except cobalt, the moment is greatest when the field is applied perpendicular to the plane of the film, similar to the anisotropic ferromagnetism recently discovered in unsubstituted $HfO_2$ [22]. In addition, some of the ZnO films on R-cut sapphire show an enormous in-plane anisotropy, with two- and four-fold symmetry. The moment per cobalt for the $Zn_{0.95}Co_{0.05}O$ film is 1.9 $\mu_B$ when the field is applied perpendicular to the plane of the film, but it varies from 2.6 to 0.8 $\mu_B$ in the plane, with the largest value measured with the field parallel to the edge of the substrate, and the smallest along a diagonal which coincides with the *c*-axis of ZnO films. The ZnO *c*-axis lies parallel to the substrate [12$\bar{1}$] axis [23].

More data for the cobalt-containing ZnO are shown in Fig. 3. The samples with 5 at% Co prepared at different oxygen pressures are *n*-type conductors when the oxygen pressure is less than 0.1 mbar, and insulators ($\rho > 10$ Ω m) at greater pressure. Heating the 1 mbar film at 700 °C in vacuum leads to an increase in moment. Carriers in ZnO are associated with oxygen vacancies or interstitial zinc atoms [24,25]. The data in Fig. 3a show that ferromagnetism is destroyed as the carriers and their associated defects are eliminated.

Figure 3b shows how the average cobalt moment in films prepared at $10^{-4}$ mbar varies with cobalt concentration. The low-concentration limit is $m \approx 5.5$ $\mu_B$, compared to the spin-only moment $m_{spin} = 3$ $\mu_B$ of high-spin Co$^{2+}$ ($e^4 t_2^3$). The fall-off in moment with increasing *x* can be rationalized in terms of a random distribution of cobalt ions over the cation sites in the wurtzite lattice. Isolated ions contribute the full moment *m*, pairs and most groups of four are antiferromagnetically coupled and make no net contribution, triplets contribute *m*/3. Large antiferromagnetically-coupled clusters of *N* atoms will make a contribution $m/N^{1/2}$. The model reproduces the trend in Fig 3b, provided the zinc distribution is weighted in favor of zinc nearest-neighbor pairs.

The variation of magnetic moment across the series shown in Fig.1 is exactly as expected from the spin-split impurity band model [18], which can account for ferromagnetism in high-*k* dielectrics doped with a few percent of transition-metal ions. For the light 3*d* elements, the 3*d*↑ states lie high in the 2*p*(O) – 4*s*(Zn) gap, overlapping the donor impurity band which is spin-split as shown in Fig. 4. In the middle of the series, there is no overlap with the 3$d$ levels and exchange is weak, but towards the end of the series the 3$d\downarrow$ states overlap the impurity band, which then has the opposite spin splitting for the same occupancy. High Curie temperatures are found whenever unoccupied 3$d$ states overlap the impurity band, but not otherwise. In ZnO films, the likely origin of the donor impurity band is lattice defects such as oxygen vacancies, which have more than one trapped electron (F$^0$ centres) [24,25].

The huge anisotropy of the magnetization of the oriented films and the large cobalt moment at low concentrations are very unusual. Giant moments were first reported by Ogale et al. for SnO$_2$ doped with Co [26], and later confirmed for both Co and Mn doping [27]. An explanation of why the cobalt orbital moment might be largely unquenched by the crystal field in ZnO is the relatively small ligand field produced by tetrahedral oxygen coordination [28] and the screening of the more distant contributions to the crystal field by the polarizable lattice. The full unquenched moment of Co$^{2+}$ is 6 $\mu_B$ (J = 9/2, g = 4/3). A negative contribution to the spin moment from the impurity-band electrons arises when it is more than half-full. It has also been suggested that cation vacancies may lead to the appearance of magnetic moments of order 0.3 $\mu_B$ on surrounding oxygen anions, which then order ferromagnetically [29]. Anion vacancies can behave similarly [30]. The data indicate an anisotropic contribution to the magnetization beyond that of the 3$d$ cation because the sign of the anisotropy does not follow the quarter-shell rule, changing with the quadrupole moment of the 3$d$ shell. The anisotropy may arise from $d$ character acquired by the impurity band, or from oriented magnetic point defects.

The valence of the substitutional 3*d* ions changes at some point along the series. Cobalt is divalent, titanium is trivalent. Iron can have either valence depending on the preparation conditions. Divalent cations simply replace zinc, but trivalent ones also act as electron dopants. However there was no obvious correlation between conductivity and cation valence (Ti, Cr and Fe are insulating, the others are conducting) or conductivity and magnetism. The electrons in the impurity band will be localized by the influence of electronic correlations and potential fluctuations [31] associated with the 3*d* cations. Localization does not preclude ferromagnetic coupling, providing the localization length is not much shorter than $\gamma a_0$, where $a_0$ is the Bohr radius.

In conclusion, the high-temperature ferromagnetism in doped zinc oxide films is an intriguing physical phenomenon, which differs from any previously-known ferromagnetism or ferrimagnetism in oxides; it appears far below the percolation threshold and is highly anisotropic in a way that does not reflect the quadrupole moment of the 3*d* charge distribution. We find a connection between the electronic structure of $(Zn_{0.95}M_{0.05})O$ and its ferromagnetic properties, which is understandable in terms of the spin-split donor impurity band model. Secondary phases cannot be the explanation of the ferromagnetism, especially when M = Sc or Ti. Electronic structure calculations are needed to define the role of defects and evaluate the *d* character acquired by the impurity-band defect states on hybridization with the M 3*d* states, in order to elucidate the nature and extent of the orbital moment. Predictions regarding the sign of the spin splitting at the bottom of the conduction band may be checked experimentally by measurements of magnetic dichroism. Further challenges are to tune the ferromagnetic properties of zinc oxide by modifying the donor concentration by compensation, or by the use of a gate electrode.

Acknowledgements:

This work was supported by Science Foundation Ireland and the Higher Education Authority of Ireland under PRTLI. We are grateful to John Donegan for helpful discussions of the optical spectra, to Cian Cullinan for the lattice simulations and to Lucio Dorneles for the substrate orientation.

Figure Captions

Figure 1. Magnetic moment of $(Zn_{0.95}M_{0.05})O$ films, M = Sc, Ti…..Cu, Zn, measured at room temperature with the field applied perpendicular to the film plane, The moment is expressed as $\mu_B$/M. The trend measured at 5 K is similar.

Figure 2. Magnetization curves of a $(Zn_{0.95}Co_{0.05})O$ film with the magnetic field applied perpendicular to the plane of the film, or in-plane in different directions. The in-plane anisotropy is shown in the insert. Data are corrected for the diamagnetism of the substrate, $-0.20 \; 10^{-12} \; m^3$.

Figure 3, a) Magnetic moment of $(Zn_{0.95}Co_{0.05})O$ measured at room temperature for films prepared at different oxygen pressures (the open symbol is for a film after annealing at 700 °C in vacuum) b) Magnetic moment of $(Zn_{1-x}Co_x)O$ measured at room temperature for films prepared at $10^{-4}$ mbar for different *x*. The solid line is based on a weighted random distribution of cobalt ions over the cation sites of the wurtzite lattice, with strong antiferromagnetic coupling of nearest-neighbour cobalt cations.

Figure 4. Schematic density of states for a) M = Ti, b) M = Mn and c) M = Co. The Fermi level lies in a spin-split donor impurity band.

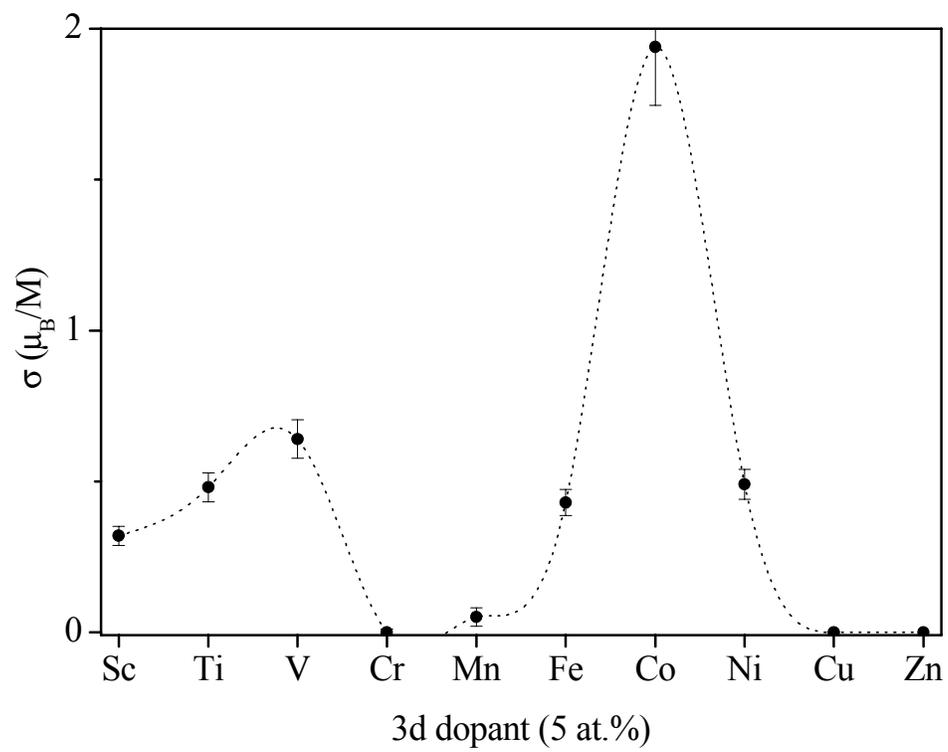

**Figure 1**

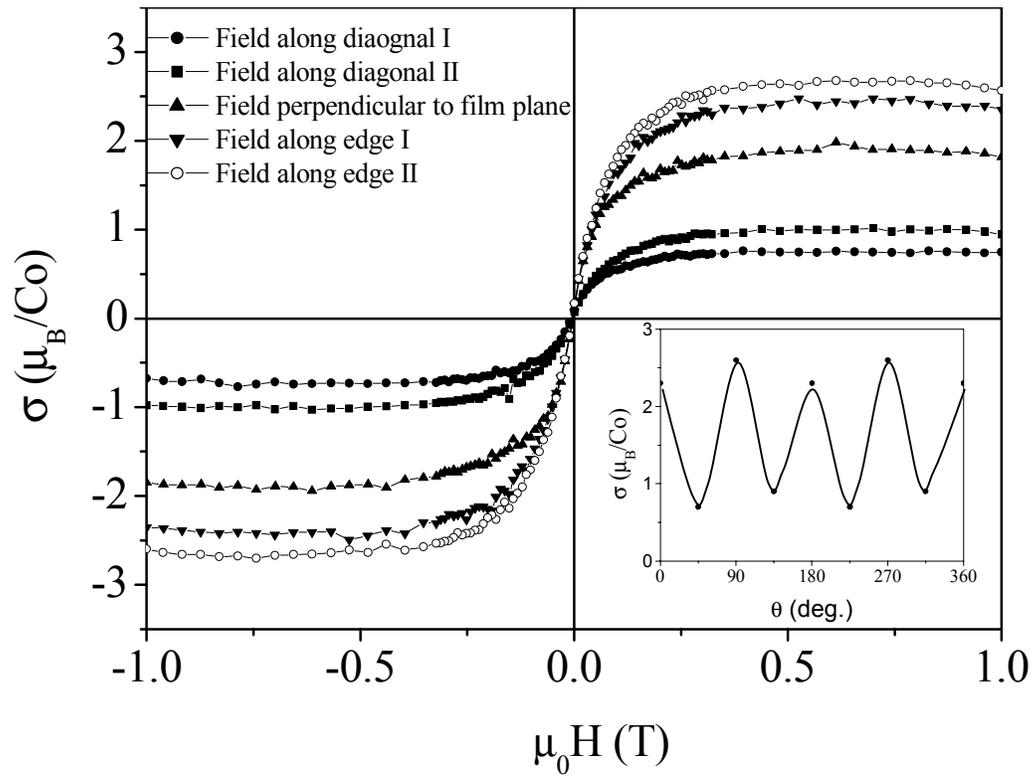

Figure 2

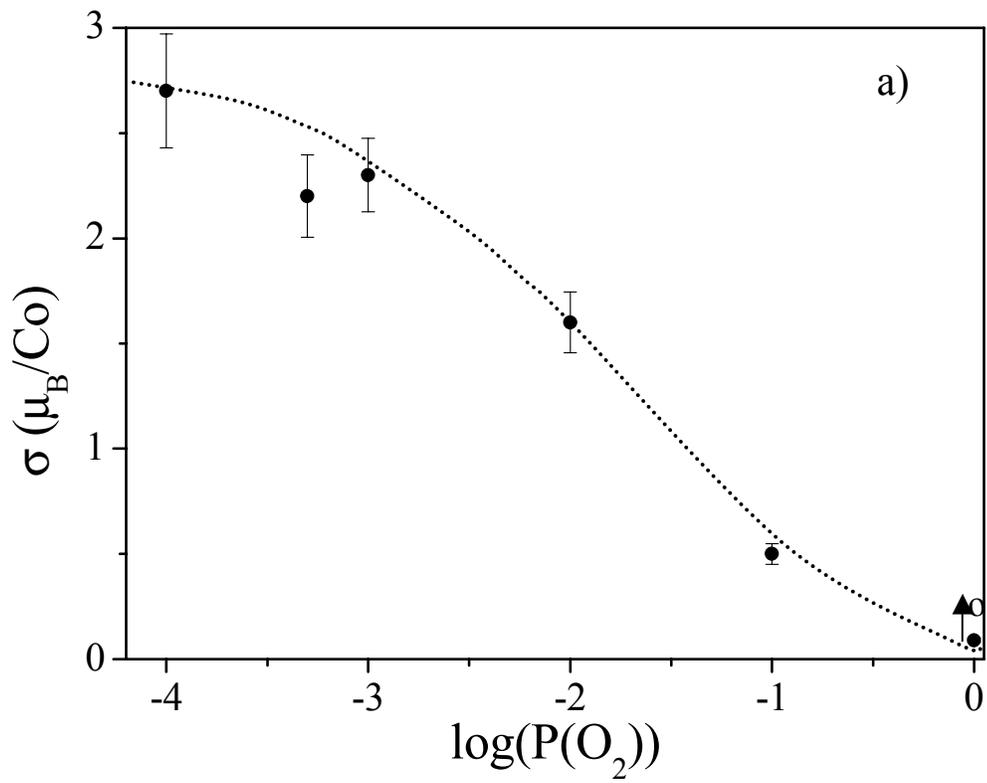
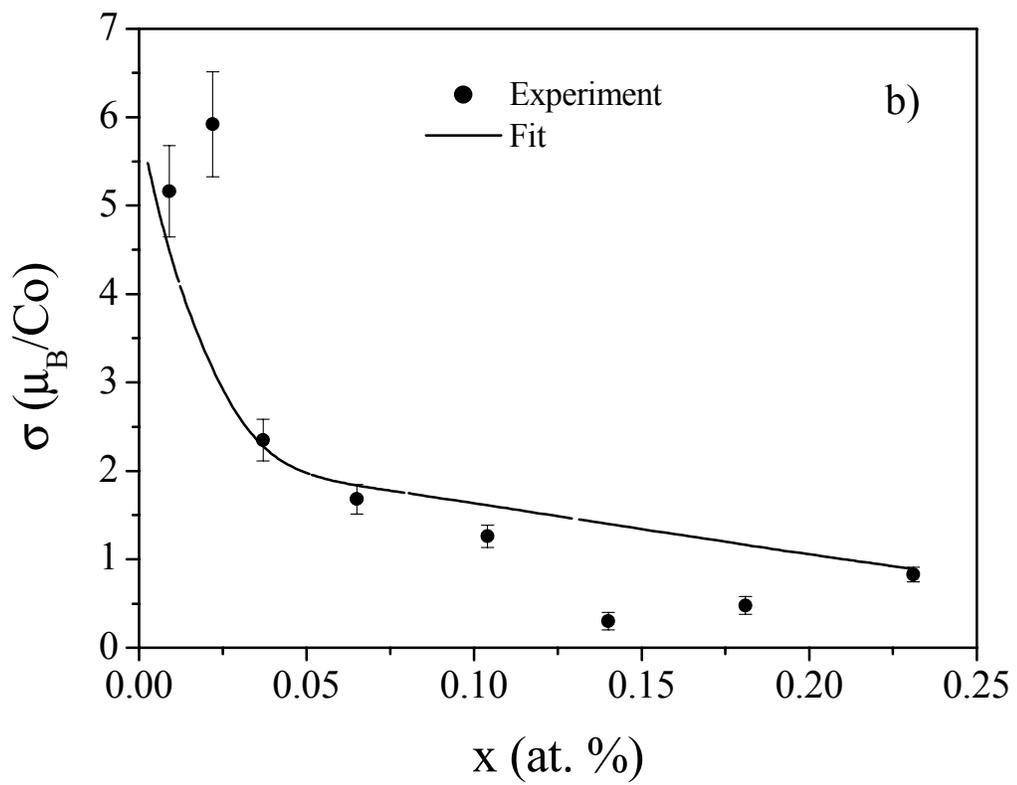

Figure 3

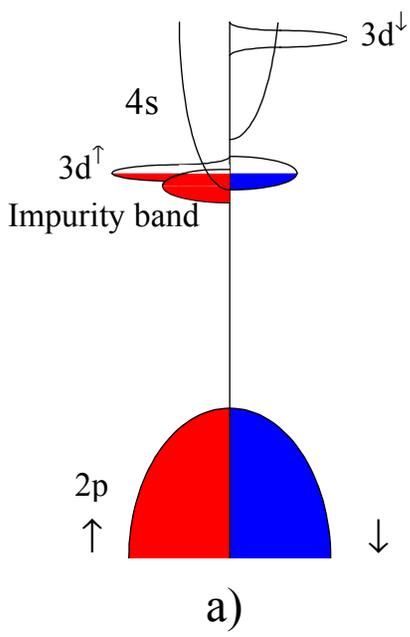 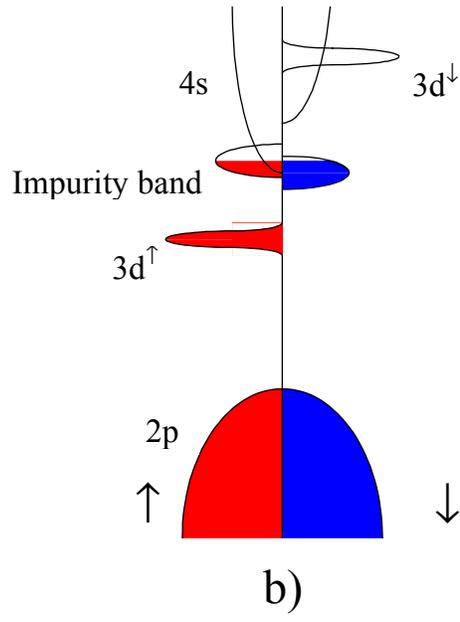 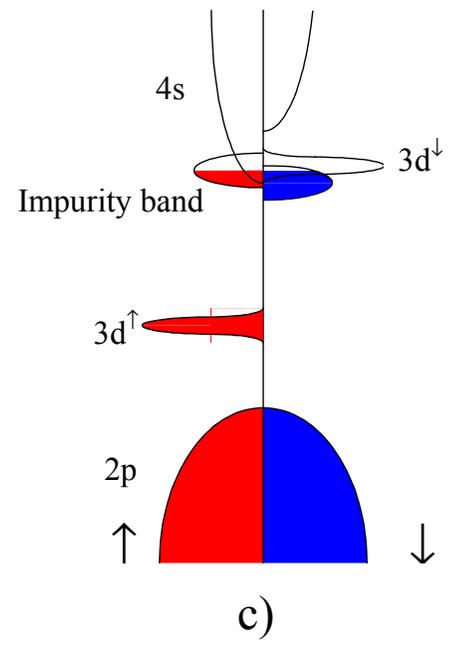

Figure 4